\documentclass[aps,pra,showpacs,reprint,superscriptaddress,superscriptaddress,twocolumn]{revtex4-1}  
\usepackage{amsmath}
\usepackage{graphicx}  
\usepackage{dcolumn}   
\usepackage{bm}        
\usepackage{amssymb}   
\usepackage{multirow}
\usepackage{hyperref}
\usepackage{subfigure}
\usepackage{epstopdf}
\begin{document}
\preprint{APS}

\title{Analytical solutions and criteria for the quantum discord of two-qubit X-states}



\author{A.Maldonado-Trapp}
\affiliation{Departamento de F\'{\i}sica, Universidad de Concepci\'{o}n, 160-C Concepci\'{o}n, Chile}
\author{Anzi Hu}
\affiliation{Department of Physics, American University, Washington, DC 20016, USA}
\author{Luis Roa}
\affiliation{Departamento de F\'{\i}sica, Universidad de Concepci\'{o}n, 160-C Concepci\'{o}n, Chile}
\date{\today}

\begin{abstract}
Except for a few special states, computing quantum discord remains  a complicated optimization process.
In this paper we present analytical solutions for computing quantum discord of the most general class of $X$-states and the criteria for each analytical solution to be valid.
We discuss parameter regions that correspond to different analytical solutions and explain the underlying reasons for such structure to exist. We apply our formalism to study both arbitrary $X$-states and $X$-states with certain symmetries. 
We find that our analytical formalism is in excellent agreement with numerical calculation of quantum discord in both cases.
\end{abstract}



\maketitle

\section{Introduction}
Quantum discord \cite{OllivierVedral,Review} as a measure of the quantumness of correlations has lead to revaluation of the advantage of quantum systems which are affected by decoherence.
Numerous are the quantum protocols which can be performed with quantum 
discord \cite{DQC1,nonlocality,noentanglement1,Discrimination,Overlap,statepreparation} instead of entanglement \cite{Wootters}.
As a measure of non-classical correlation, quantum discord is also recognized as a fundamental property of quantum systems \cite{Adesso,Piranola,Unified,Auyuanet,Gaussian}.
Although the definition of quantum discord is conceptually clear, computing it has been difficult since it requires the optimization over all possible measurements on one qubit.
The most general class of quantum states for which quantum discord can be studied analytically has been shown to be the $X$ states \cite{NP}.
$X$-states\cite{Rau2009} are encountered frequently in different contexts of physics\cite{Fedorov,spinmodels,Werlang,Sarandy,XYchain,swapping,Xstates}.
Although there have been a few methods for finding  analytical results for two-qubit $X$-states \cite{Luo,PrimeroX,Fanchini,Analyticalprogress,Optimalmeasurements,Proyective,Quesada} they are not always correct.
Extensive numerical studies have shown that the two known analytical formulas are not always valid and errors exist in the calculation of quantum discord with such formulas \cite{Yichen}.

In this paper we consider the most general form of $X$-states and we find that the quantum discord for the majority of the states can be calculated with two closed form formulas.
We also determine analytically the criteria for these formulas to be valid and the solution for the rare cases where the two formulas are not valid. 
This provides a way to calculate the quantum discord for any $X$-states without performing the optimization process.
The article is arranged as follows:
In Sec. \ref{sec 2} we introduce the definition of quantum discord \cite{OllivierVedral}.
In Sec. \ref{sec 3} we discuss the derivation of our formalism.
In Sec. \ref{sec 4} we discuss examples and applications to various $X$-states and the comparison to numerical calculation.
Finally, in the last section we summarise our principal results and future directions.
\section{Quantum Discord  \label{sec 2}}

Quantum discord arises from the difference between two classically equivalent definitions for mutual information that are related through Bayes' 
theorem \cite{OllivierVedral}.
Firstly, the quantum mutual information, $I(\rho_{AB})$, is the total shared information between two quantum subsystems $A$ and $B$ being in a state $\rho_{AB}$, specifically \cite{Thomas}
\begin{equation}
I(\rho_{AB})=S(\rho_A)+S(\rho_B)-S(\rho_{AB}), \label{qmi1}
\end{equation}
where $S(\rho_A)$, $S(\rho_B)$, and $S(\rho_{AB})$ are the von Neumman entropies \cite{Nielsen} of the reduced states $\rho_A=Tr_B(\rho_{AB})$, $\rho_B=Tr_A(\rho_{AB})$, and the total state $\rho_{AB}$ respectively.
Secondly, the quantum mutual information, $J\left(B|A\right)$, can be defined based on the quantum equivalence of Bayes' theorem.
It represents the information gained about the subsystem  $B$ after measuring the set $\Pi_a$ on subsystem $A$ and it is given by
\begin{equation}
J\left(B|A\right)=S(\rho_B)-\min_{\{\Pi_{a}\}}S\left(\rho_{B|A}\right).  \label{j}
\end{equation}
Here $S\left(\rho_{B|A}\right)=\sum_a p_a S\left(\rho_{B|a}\right)$ is the average conditional entropy of system $B$, where $p_a=Tr(\Pi _{a} \rho_{AB}\Pi _{a}^\dag)$ is the probability to obtain the outcome $a$ and $\rho_{B|a}=\Pi _{a}\rho_{AB}\Pi _{a}^\dag/p_a$ is the corresponding reduced density for subsystem $B$.
The minimization of conditional entropy is performed over all the possible positive operator valued measure (P.O.V.M.) \cite{POVM}.
Quantum discord is defined as the discrepancy between $I(\rho_{AB})$ and $J\left(B|A\right)$,
\begin{eqnarray}
D(B|A)&\equiv &I(\rho_{AB})-J(B|A), \nonumber\\
&=&\min_{\{\Pi_{a}\}}S\left(\rho_{B|A}\right)-S(\rho_{AB})+S(\rho_A) \label{discorddefinition}.
\end{eqnarray}
Quantum discord can be understood as a measurement of the non-Bayesianism of the state $\rho_{AB}$, or the amount of shared information that is not accessible by local measurements.
It should be noticed that the information about $B$ given a measurement on $A$ is not necessary the same as the information of $A$ given a measurement on $B$ and therefore $D(B|A)\neq D(A|B)$.
The definition given in Eq.(\ref{discorddefinition}) shows the challenge of calculating discord is to find the optimal measurement that minimizes the average conditional entropy $S\left(\rho_{B|A}\right)$. 
Determining the optimal measurement is the focus of our discussion in the next section.

\section{Formulism  \label{sec 3}}

Let us consider two qubits $A$ and $B$ in $X$-state $\rho_{X}$, which in the logic basis $\{\left\vert 00\right\rangle$, $\left\vert
01\right\rangle$, $\left\vert 10\right\rangle$, $\left\vert 11\right\rangle\}$ has the following matrix representation
\begin{equation}\label{Xstate}
\rho _{X}\equiv\left(
\begin{array}{cccc}
a & 0 & 0 & w \\
0 & b & z & 0 \\
0 & z & c & 0 \\
w & 0 & 0 & d
\end{array}\right).
\end{equation}%
Here the diagonal elements are complementary probabilities, $a+b+c+d=1$, and the positivity of the density matrix requires $|w|\leq\sqrt{ad}$ and $|z|\leq\sqrt{bc}$.
We consider $w$ and $z$ real and positive numbers since their phases can be removed by local unitary operation \cite{Review} and do not change the quantum correlations between $A$ and $B$.
For arbitrary two qubit states of rank 2, the P.O.V.M that minimizes the conditional entropy is the projective von Neumann measurement\cite{Proyective}. For states of rank 3 and 4 F. Galve et al. \cite{Proyective} showed that projective measurements are almost sufficient, except for states that appear with probability $10^{-2}$ and in which case the deviations are on average of the order of $10^{-6}$.
The projective measurement bases are characterized by the angles $\theta$ and $\phi$ of the Bloch sphere, as follows
\begin{eqnarray}
|+\rangle&=&\cos(\theta/2)|0\rangle+e^{i\phi}\sin(\theta/2)|1\rangle,  \label{vN1} \\
|-\rangle&=&\sin(\theta/2)|0\rangle-e^{i\phi}\cos(\theta/2)|1\rangle, \label{vN2}
\end{eqnarray}
with $\theta\in[0,\pi]$ and $\phi\in[0,2\pi]$.
We consider measuring A with the above measurement bases, $\Pi_{\pm}=|\pm\rangle\langle\pm|$. The corresponding outcome probabilities $p_{\pm}$ are
\begin{equation}
p_{\pm}=\frac{1}{2}(1\pm A_2\cos\theta),
\end{equation}
and the average conditional entropy of $B$ as a function of $\theta$ and $\phi$, $F(\theta,\phi)$, is written as
\begin{eqnarray}
F(\theta,\phi)&=&p_{+}S(\rho_{B|+})+p_{-}S(\rho_{B|-}),  \nonumber\\
&=&-\frac{1}{2}\left( T_{+} +T_{-} + Q_{+} +Q_{-}\right), \label{FdeTyPh}
\end{eqnarray}
with
\begin{eqnarray}
T_{\pm }\left( \theta ,\phi \right)  &=&\left( p_{+}\pm R_{+}\right) \log_{2}\frac{p_{+}\pm R_{+}}{2p_{+}} , \\
Q_{\pm }\left( \theta ,\phi \right)  &=&\left( p_{-}\pm R_{-}\right) \log_{2}\frac{p_{-}\pm R_{-}}{2p_{-}}.
\end{eqnarray}
To simplify the analysis, we define following functions
\begin{eqnarray}\label{B2}
B^{2}&=&w^{2}+z^{2}+2wz\cos(2\phi), \\
G_{\pm}&=&\frac{A_{3}\pm A_{1}\cos\theta}{2}, \\
R_{\pm }&=&\sqrt{G_{\pm }^{2}+B^{2}\sin^2\theta},
\end{eqnarray}
and constants
\begin{eqnarray}
A_{1}&=&a-b-c+d,\\ \nonumber
A_{2}&=&a+b-c-d,\\ \label{aes}
A_{3}&=&a-b+c-d. \nonumber
\end{eqnarray}
It is worth noting that $F(\theta,\phi)=F(\theta,2\pi-\phi)$ and $F(\theta,\phi)=F(\pi-\theta,\phi)$, i.e. $F$ is symmetric with respect to $\phi=\pi$ and $\theta=\pi/2$.
It is therefore sufficient to consider $\phi\in[0,\pi[$ and $\theta\in[0,\pi/2]$.
Furthermore, the only $\phi$ dependence of $F$ is in the term $wz\cos(2\phi)$ of $B^2$ in Eq.(\ref{B2}) and $F$ becomes independent of $\phi$ when $w$ or $z$ is zero.
We also note $A_1$, $A_2$ and $A_3$ are the expectation values of the $z$-component of the Pauli vector, $A_1=\langle\sigma_{z}^A\sigma_{z}^B\rangle$, $A_2=\langle\sigma_{z}^A\rangle$ and $A_3=\langle\sigma_{z}^B\rangle$
 \cite{Nielsen}.
The minimum of $F(\theta,\phi)$ can be found by using the extreme value theorem \cite{Calculo}.
The critical points $\textbf{c}=(\theta_c, \phi_c)$ are determined by solving the first partial derivatives equations $\partial F/\partial \theta=0$ and $\partial F/\partial \phi=0$ and whether the critical point $\textbf{c}$ corresponds to a minimum is determined by the second partial derivatives test. i.e.
the second partial derivatives and the Hessian matrix determinant have to be simultaneously positive at $\textbf{c}$
\begin{equation}
\left.\frac{\partial^2F}{\partial\theta^2}\right\vert_{\textbf{c}}>0,\quad \left.\frac{\partial^2F}{\partial\phi^2}\right\vert_{\textbf{c}}>0,\quad
\det\left.\left(
\begin{array}{cc}
\frac{\partial ^{2}F}{\partial \theta ^{2}} & \frac{\partial ^{2}F}{\partial
\theta \partial \phi } \\
\frac{\partial ^{2}F}{\partial \theta \partial \phi } & \frac{\partial ^{2}F
}{\partial \phi ^{2}}%
\end{array}
\right)\right\vert_{\textbf{c}}>0. \nonumber
\end{equation}
Otherwise the critical point \textbf{c} can correspond to a local maximum, a saddle point, or if Hessian matrix determinant is zero, higher order test must be used to determine the nature of the critical point.\\

We find the first partial derivatives of $F$ as
\begin{eqnarray}
\frac{\partial F}{\partial \phi } &=&-wz\sin ^{2}\theta\sin(2\phi)C_{\phi } \label{cphi},\\
\frac{\partial F}{\partial \theta } &=&-\frac{\sin \theta }{4}C_{\theta }, \label{ctheta}
\end{eqnarray}%
with
\begin{equation}
C_{\phi}=\frac{1}{R_{-}}\log _{2}\frac{p_{-}+R_{-}}{p_{-}-R_{-}}+\frac{1}{R_{+}}\log _{2}\frac{p_{+}+R_{+}}{p_{+}-R_{+}},
\end{equation}%
and
\begin{widetext} 
\begin{equation}
C_{\theta}\!=\!\frac{A_{1}G_{-}+2B^{2}\cos\theta}{R_-}\log_{2}\frac{p_{-}+R_{-}}{p_{-}-R_{-}} -\frac{A_{1}G_{+}-2B^{2}\cos\theta}{R_+}\log_{2}\frac{p_{+}+R_{+}}{p_{+}-R_{+}}+A_{2}\log_{2}\frac{p_{+}^{2}\left(p_{-}^{2}-R_{-}^{2}\right) }{p_{-}^{2}\left( p_{+}^{2}-R_{+}^{2}\right)}.
\end{equation}
\end{widetext}
From these expressions we note several key features about the critical points:
\begin{itemize}
  \item Because  $0 < R_\pm < p_{\pm}$, $C_\phi$ is always positive and the partial derivative of $F$ with regard to $\phi$ in  Eq.(\ref{cphi}) is zero   only when $\sin (2\phi)$ is zero for any $\theta$, i.e $\phi_c=0$, $\pi/2$, and when $\theta=0$ for any $\phi$. We note that if $\theta$ is zero then the minimization does not depend on $\phi$.
  At these $\phi_c$ the second derivative $\partial ^2F/\partial \phi^2$ becomes
  \begin{equation}
  \left. \frac{\partial^{2}F}{\partial \phi ^{2}}\right\vert_{(\theta,0) } = 2wzC_{\phi=0}\sin^{2}\theta >0
  \end{equation}  
  $\forall$ $\theta\neq0$ and
   \begin{equation}
  \left. \frac{\partial^{2}F}{\partial \phi ^{2}}\right\vert_{\left(\theta,\frac{\pi}{2}\right) } =-2wz C_{\phi=\frac{\pi }{2} }\sin^{2}\theta<0
  \end{equation}
$\forall$ $\theta\neq0$.\\
	Since for any $\theta$ the second derivative  $\partial ^2F/\partial \phi^2$ is always negative when $\phi=\pi/2$ the minimization problem becomes one 		variable minimization problem where $\phi_c=0$. 
	As $F(\theta,0)$ becomes a one variable function is not necessary consider the Hessian determinant it is sufficient study the behavour of $\left.		\frac{\partial^2F}{\partial\theta^2}\right\vert_{\textbf{c}}$.
  \item The derivative Eq.(\ref{ctheta}) is zero at either $\sin\theta=0$ or $C_{\theta}=0$. 
  The former equation provides $\theta_c=0$. 
  The equation $C_{\theta}=0$ has the obvious root $\theta_c=\pi/2$ and a special root $\theta_e$. 
  This special root depends on the density matrix elements and only appears in some very particular states, such as shown in Fig.\ref{1}$(b)$.
  The angle $\theta_e$ is not considered in previous analytical studies \cite{Rau2009,PrimeroX,Fanchini}. 
  It corresponds to the ``error cases'' reported in numerical calculations \cite{Yichen}. 
\end{itemize}
Based on these features, the optimization problem reduces to studying the second derivative of $F(\theta,0)$ with respect to $\theta$ evaluated at the critical angles $\theta_c=0$, $\pi/2$, and $\theta_e$. 
We notice that the measurement bases become $\sigma_z$ for $\theta_c=0$ and $\sigma_x$ for $\theta_c=\pi/2$. 
The value for $\theta_e$ remains dependent on density matrix elements and we denote this measurement as $\sigma_e$.

\begin{table}[t]
\centering
\caption{Criterion for different optimal measurement ($M$) in terms of Pauli matrices and the solution to $\min F(\theta,\phi)$. 
The measurements of $\sigma_z$ and $\sigma_x$ are the predominant cases for an arbitrary X-state. 
They include more than 99\% of the parameter space. 
For the case where the measurement is listed as $any$, $F(\theta,\phi)$ is the same for any projective measurement.
For the case  where the measurement is listed as $\sigma_e$, it is necessary find $\theta_e$ from $C_{\theta}=0$ solving for every set of density matrix elements.
For case of $\sigma_?$, $\min F(\theta,\phi)$ is the minimum of $F(0,0)$ and $F\left(\frac{\pi}{2},0\right)$.}
\begin{tabular}{|c|c|c|}
\hline
$M$ & Condition & $\min F\left( \theta ,\phi \right) $ \\ \hline \hline
$any$ & $C_{0}= 0 \vee C_{+}=0$ & $F(0,0)$ \\ \hline
$\sigma_z$ & $C_{0}<0 \wedge C_{+}>0$ & $F(0,0)$ \\ \hline
$\sigma_x$ & $C_{0}>0 \wedge C_{+}<0$ & $F(\frac{\pi}{2},0)$ \\ \hline
$\sigma_e$ & $C_{0}>0 \wedge C_{+}>0$ & $F(\theta_{e},0)$ \\ \hline
$\sigma_?$ & $C_{0}<0 \wedge C_{+}<0$ & $\min\left\{F(0,0),F(\frac{\pi}{2},0)\right\}$ \\ \hline
\end{tabular}
\label{tabla1}
\end{table}

The second derivative evaluated in the three $\theta_c$ depends on the behaviour of two quantities $C_{0}$ and $C_+$ given by
\begin{widetext}
\begin{equation}
C_{0} =A_{2}\log _{2}\frac{cd\left( 1+A_{2}\right) ^{2}}{ab\left(
1-A_{2}\right) ^{2}}-A_{1}\log _{2}\frac{ad}{bc}+2\left( w+z\right)^2 \left( \frac{1}{a-b}\log _{2}\frac{a}{b}+\frac{1}{c-d}%
\log _{2}\frac{c}{d}\right) 
\end{equation}%
and
\begin{equation}
C_{+} =4r\left( A_{1}A_{3}-4A_{2}r^{2}\right) ^{2}-4\left( w+z\right) ^{2}\left( 1-4r^{2}\right) \left( 4r^{2}-A_{1}\right)
\ln \frac{1+2r}{1-2r}
\end{equation}
\end{widetext}
where $r=\sqrt{A_3^2+4\left(w+z\right)^{2}}/2$.
The behaviour of $C_0$ and $C_+$ determines which of $F(0,0)$, $F(\pi/2,0)$, and $F(\theta_e,0)$ is a minimum. 
The relationship is summarized in Table \ref{tabla1}. 
It is worth noting that for more than 99\% parameter space  \cite{Yichen}, $F(\theta,\phi)$ is minimized either at $F(0,0)$ or $F(\pi/2,0)$, which can be calculated as follows,
\begin{equation}\begin{small}\label{F0}
F(0,0)=-a\log_{2}\frac{a}{a\!+\!b}-b\log_{2}\frac{b}{a\!+\!b}-c\log_{2}\frac{c}{c\!+\!d}-d\log_{2}\frac{d}{c\!+\!d},
\end{small}
\end{equation}%
and
\begin{small}
\begin{eqnarray}\label{F2}\nonumber
F\!\left(\frac{\pi}{2},0\right) &\! =\! & -\! \frac{1\!+\!\sqrt{\!A_3^2\!+\!4(w\!+\! z)^2}}{2}\log_2\frac{1\!+\!\sqrt{\!A_3^2\!+\!4(w\!+\! z)^2}}{2}\\
&\! &\!-\!\frac{1\!-\!\sqrt{\!A_3^{2}\!+\!4(w\!+\! z)^{2}}}{2}\log_{2}\frac{1\!-\!\sqrt{\!A_3^{2}\!+\!4(w\!+\! z)^{2}}}{2}.
\end{eqnarray}
\end{small}

The analytical solution also agrees with the particular expressions obtained in Ref.\cite{Fanchini} for symmetric X-state, i.e with $b=c$ in the density matrix of Eq.(\ref{Xstate}). 
We find that there are three possible minima for an arbitrary $X$-state and we find for the first time the conditions for special cases $\sigma_e$ and $\sigma_?$. 
In the following section, we discuss examples where different measurements are necessary and use our formula to calculate the quantum discord. 

\section{Results \label{sec 4}}
\subsection{General $X$-states  \label{sec 4a}}
\begin{figure}[hbtp]
\centering
\includegraphics[scale=0.6]{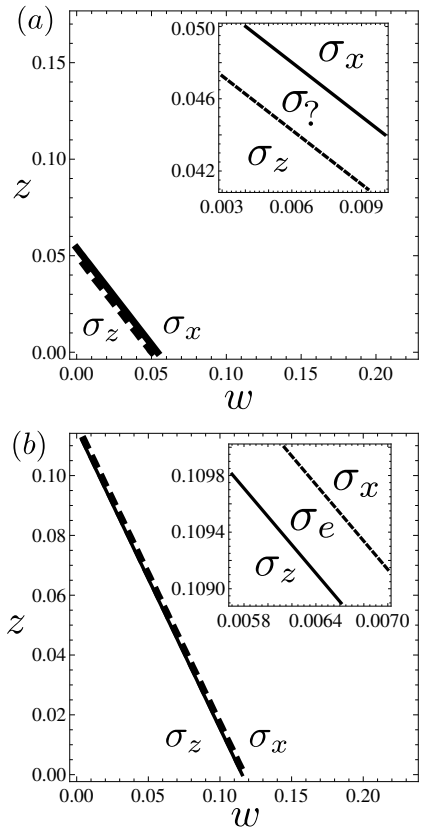}
\caption{Regions where the different cases arise as function of $w$ and $z$, for two particular $X$-states: ($a$) the diagonal elements are $a=0.5$, $b=0.3$, $c=0.1$ and $d=0.1$, ($b$) the diagonal elements are $a=0.0783$, $b=0.1250$, $c=0.1$ and $d=0.6967$. The solid and dashed line correspond to $C_0=0$ and $C_+=0$, respectively.}
\label{1}
\end{figure}
First we consider two examples of general $X$-states where the rare measurement cases exist.
In Figure \ref{1}$(a)$, we show an example where the optimal measurements changes from $\sigma_z$, $\sigma_?$ and $\sigma_x$ as $w$ and $z$ changes with the diagonal density matrix elements fixed.
For the parameter space where $w$ and $z$ go from zero to approximately $0.0502681$ the optimal measure is always $\sigma_z$, as indicated by the dashed line.
For the parameter space where $w$ and $z$ go from approximately $0.0540158$ and above, the optimal measure is $\sigma_x$, as indicated by the solid line in the plot.
In the extremely small region (shown as the insert plot), where $w$ and $z$ go from approximately $0.0502681 $ to $0.0540158$, the measurement is denoted as $\sigma_?$.
This is the region where both $C_0$ and $C_+$ are negative and the optimal measurement has to be determined by calculating $F(0,0)$ and $F(\pi/2,0)$. 
In Figure \ref{1}$(b)$, we show an example where the optimal measurements changes from $\sigma_z$, $\sigma_e$ and $\sigma_x$ as $w$ and $z$ changes with the diagonal density matrix elements fixed. 
For $w$ from zero to approximately $0.233563$ and $z$ from zero to approximately $0.111803$ the optimal measure is always $\sigma_z$, as indicated by the solid line. 
When $w$ is greater than approximately $0.116136$ and $z$ is greater than approximately $0.111803$, the optimal measure is $\sigma_x$, as indicated b the dashed line in the plot. 
In the small region between the dashed and solid lines (shown in the insert plot), neither $F(0,0)$ nor $F(\pi/2,0)$ is the minimum.
In this region the optimal measurement is given by the special measurement $\sigma_e$. We also note that our criterion reproduces the results obtained in  references \cite{Yichen} and \cite{Optimalmeasurements}.

\begin{figure}[hbtp]
\centering
\includegraphics[scale=0.74]{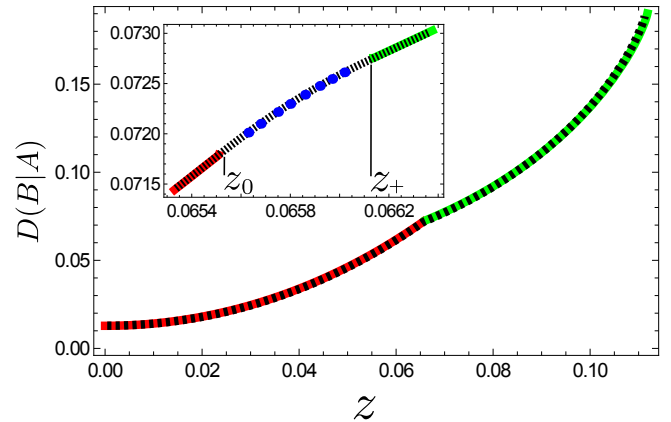}
\caption{Quantum discord, $D(B|A)$, as function of $z$ for $a=0.0783$, $b=0.1250$, $c=0.1$, $d=0.6967$ and $w=0.05$. 
Black dashed line correspond to numerical solution of Eq.(\ref{discorddefinition}) solved for every value $z$.
Red and green lines are our analytical calculation based on Eq.(\ref{F0}) and Eq.(\ref{F2}). 
The optimal measurement for the red line is $\sigma_z$. 
The optimal measurement for the green line is $\sigma_x$. 
The blue dots are the result from the special measurement $\sigma_e$, which is determined by the roots of $C_\theta=0$ for a given $z$ value.
%
%
%
}
\label{discord}
\end{figure}

\begin{figure}[hbtp]
\centering
\includegraphics[scale=0.6]{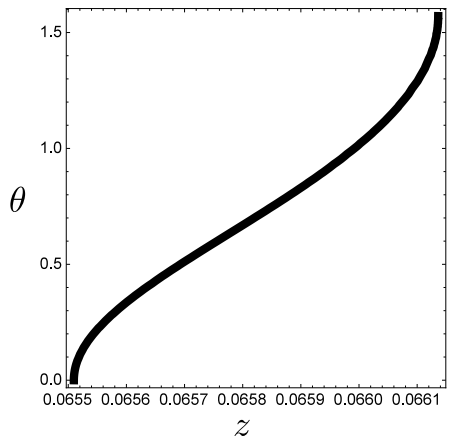}
\caption{ Special angle $\theta_e$ of the measurement $\sigma_e$ as a function of $z$ in the special measurement region of the state given in Fig. 2. 
In this region, the special changes continuously from zero to $\pi/2$. 
}
\label{theta}
\end{figure}
Figure (\ref{discord}) shows a comparison between numerical\cite{note} and analytical results for quantum discord. 
Here discord is calculated for the density matrices discussed in Fig.\ref{1}$(b)$ with $w=0.05$, and the optimal measurement can be either $\sigma_z$, $\sigma_x$ or $\sigma_e$.
The numerical solution of discord is given by the dashed line.
When $z$ goes from zero to $z_0\approx 0.0655096$ the optimal measure is $\sigma_z$ (red line). 
When $z$ goes from $z_+\approx 0.0661362$ to its maximum value $z_{max}=\sqrt{bc}$ the optimal measure is $\sigma_x$ (green line).
When $z$ goes from $z_0$ to $z_+$ the optimal measure is $\sigma_e$ (blue dots). 
In this last case it is necessary solve $C_\theta(\theta_e,0)=0$ for every $z$. The numerical and analytical results agree.
In Fig.(\ref{theta}) we show $\theta_e$ as a function of $z$ in the special measurement regime as indicated in Fig.(\ref{discord}).
Although in Fig.(\ref{discord}) the transition between the optimal measures $\sigma_z$ and $\sigma_x$ appears sharp, Fig.(\ref{theta}) shows that for $z$ from $z_0$ to $z_+$, the special angle $\theta_e$ goes continuously from $\theta=0$ to $\theta=\frac{\pi}{2}$. 

\subsection{States with symmetries  \label{sec 4b}}
The analytical derivation in Sec. also shows that the behaviour of $F$ strongly depends on the constants $A_i$ as defined on Eq.(\ref{aes}). These constants corresponds to the expectation values of $\sigma_z$s and they become zero when certain symmetries are present. For example, a system with the spin-flip symmetry, such as the $XXZ$ model, satisfies $A_2=\langle\sigma_z^A\rangle=0$, $A_3=\langle\sigma_z^B\rangle=0$ and $A_1=\langle\sigma_z^A\sigma_z^B\rangle=4a-1$ with $0\leq a\leq1/2$. 
The most general form of the reduced density matrix of the two spins in such systems are written as
\begin{equation}\label{xxz}
\rho=
\left(
\begin{array}{cccc}
a & 0 & 0 & w \\
0 & \frac{1}{2}-a & z & 0 \\
0 & z & \frac{1}{2}-a & 0 \\
w & 0 & 0 & a%
\end{array}%
\right),
\end{equation}
with $0\leq w+z\leq1/2$. 
For such systems the only optimal measurements are $\sigma_x$ and $\sigma_z$. 
For $0\leq w+z\leq\left\vert 2a-\frac{1}{2}\right\vert$, the optimal measurement is  $\sigma_x$ and the minimum of $F(\theta,\phi)$ is 
\begin{equation}
F(0,0)=-2a\log_2(2a)-(1-2a)\log_2(1-2a). \label{Fo}
\end{equation}
For $\left\vert 2a-\frac{1}{2}\right\vert \leq w+z\leq \frac{1}{2}$, the optimal measurement is $\sigma_z$ and the minimum of $F$ is 
\begin{eqnarray}\nonumber
F\left(\frac{\pi}{2},0\right)&=&-\frac{1+2(w+z)}{2}\log_2\frac{1+2(w+z)}{2}\\
& & -\frac{1-2(w+z)}{2}\log_2\frac{1-2(w+z)}{2}.  \label{Fp}
\end{eqnarray}

\begin{figure}[hbtp]
\centering
\includegraphics[scale=0.75]{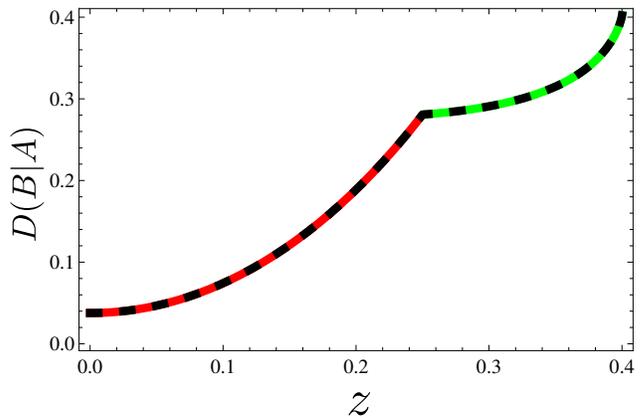}
\caption{Quantum discord for two spins in the $XXZ$ model with $a=0.1$ and $w=0$. Black dashed line correspond to numerical calculation. The red line corresponds to where the optimal measurement is $\sigma_z$ and the quantum discord is determined through calculating Eq.(\ref{Fo}). The green line corresponds to where the optimal measurement is $\sigma_x$ and the quantum discord is determined through calculating Eq.(\ref{Fp}).}
\label{xxz1}
\end{figure}

\begin{figure}[hbtp]
\centering
\includegraphics[scale=0.6]{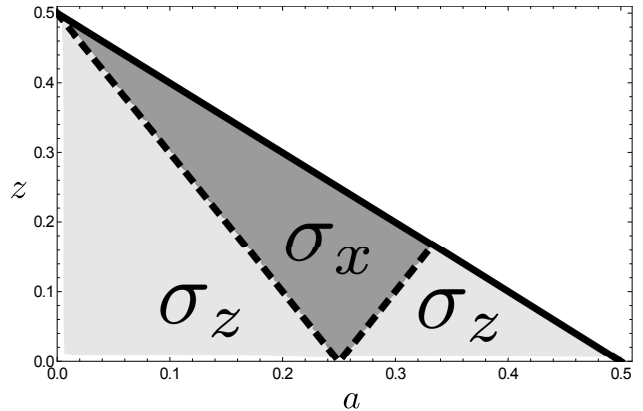}
\caption{Optimal measurements on the $z-a$ plane with $w=0$. This corresponds to the $XXZ$ model. 
Positivity of the density matrix requires that  $z\leq 1/2-a$.
The light grey area denotes the parameter region for the optimal measurement $\sigma_z$ and the dark grey area for the optimal measurement $\sigma_x$.
The dashed line corresponds to  $z=\left\vert 2a-1/2\right\vert$ and the solid line correspond to $z=1/2$. 
On the dashed and solid lines $C_0$ and $C_+$ are both zero. 
This corresponds to the first case in table (\ref{tabla1}), where $F$ is independent of the measurement angle.
 }
\label{reg}
\end{figure}

As a example we consider the state of two spins in the $XXZ$ model, which corresponds to $w=0$ in Eq.(\ref{xxz}). 
In Fig.(\ref{xxz1}) we show the typical behaviour of the quantum discord between two spins in the $XXZ$. 
We consider varying $z$ and $a=0.1$ in the density matrix.
The optimal measurement changes from $\sigma_z$ to $\sigma_x$ as $z$ increases.
On the transition point from $\sigma_z$ to $\sigma_x$, the discord is independent of the measurement angle. 
We find that numerical and analytical calculations agree perfectly for all $z$. 
In Fig.(\ref{reg}) we show the parameter regions for optimal measurements $\sigma_z$ and $\sigma_x$ for a general two-spin state in the $XXZ$ model. 
We find that when $z$ is greater than $1/2-2a$ and less than $2a-1/2$, the the optimal measurement is $\sigma_x$. 
In the rest of the parameter space the optimal measurement is $\sigma_z$. 
On the border of the parameter regions, as indicated by the dashed and solid lines, quantum discord does not depend on the measurement angle.
\section{Summary}

In this paper we analyse the quantum discord of arbitrary two-qubit $X$-states.
We describe a straightforward criterion for determining analytical solution to quantum discord and it has shown excellent agreement with numerical calculations for both arbitrary $X$-states and $X$-states with special symmetries. 
We find that the optimization of measurements to minimize the averaged conditional entropy, the key and most computationally intensive element of calculating quantum discord, is reduced to three possible cases. 
Two of them are the optimal measurements discussed in previous literature and one additional one is the "error" cases that have been observed numerically previously. 
We provide a complete framework to categorize these measurements and determine their parameter regions based on the analytical criteria we derived. 
In addition to simplifying the quantum discord calculation, our analytical expressions reveals the relationship between the optimal measurements and symmetries of the states.

Our analytical expression also has experimental implications in a sense that it provides a way to calculate the quantum discord based on the expectation value of spin components, which can be measured experimentally. 
Whether our analytical solutions provides more insight on the relationship between the behaviour of the optimal measurement and the critical phenomena, as previous numerical study suggests \cite{Dillenschneider}, remains to be further investigated.

\begin{acknowledgements}
A.M.T acknowledges the financial support of CONICYT.
\end{acknowledgements}


\begin{thebibliography}{}
\bibitem{OllivierVedral} H. Ollivier and W. H. Zurek, Phys. Rev. Lett. \textbf{88}, 017901 (2001);
L. Henderson and V. Vedral, J. Phys. A: Math. Gen. \textbf{34}, 6899 (2001).

\bibitem{Review} Kavan Modi, Aharon Brodutch, Hugo Cable, Tomasz Paterek, and Vlatko Vedral. Rev. Mod. Phys. \textbf{84}, 1655 (2012).

\bibitem{DQC1} E. Knill and R. Laflamme, Phys. Rev. Lett. \textbf{81}, 5672 (1998); A.Datta, A.Shaji, and C.M.Caves, Phys. Rev. Lett. \textbf{100}, 050502 (2008).

\bibitem{nonlocality} Charles H. Bennett, David P. DiVincenzo, Christopher A. Fuchs, Tal Mor, Eric Rains, Peter W. Shor, John A. Smolin, and William K. Wootters, Phys. Rev. A \textbf{59}, 1070 (1999).

\bibitem{noentanglement1}B. P. Lanyon, M. Barbieri, M. P. Almeida, and A. G. White, Phys. Rev. Lett. \textbf{101}, 200501 (2008).

\bibitem{Discrimination} Luis Roa, J. C. Retamal, and M. Alid-Vaccarezza, Phys. Rev. Lett. \textbf{107}, 080401 (2011).

\bibitem{Overlap}Luis Roa, A. Maldonado-Trapp, Cristian Jara-Figueroa, and Mari\'a Loreto L. de Guevara J. Phys. Soc. Jpn. \textbf{83}  044006 (2014).

\bibitem{statepreparation} Borivoje Daki\'c, Yannick Ole Lipp,	Xiaosong Ma, Martin Ringbauer, Sebastian Kropatschek, Stefanie Barz, Tomasz Paterek, Vlatko Vedral, Anton Zeilinger, Caslav Brukner et al. Nature Physics \textbf{8}, 666–670 (2012).

\bibitem{Wootters} W. K. Wootters, Phys. Rev. Lett. \textbf{80} 2245 (1998);
S. Hill and W. K. Wootters, Phys. Rev. Lett. \textbf{78} 5022 (1997).


\bibitem{Adesso} G. Adesso and A. Datta, Phys. Rev. Lett. \textbf{105}, 030501 (2010).

\bibitem{Piranola}Stefano Pirandola, Gaetana Spedalieri, Samuel L. Braunstein, Nicolas J. Cerf and Seth Lloyd, Phys. Rev. Lett. \textbf{113}, 140405 (2014)

\bibitem{Unified}Kavan Modi, Tomasz Paterek, Wonmin Son, Vlatko Vedral, and Mark Williamson, Phys. Rev. Lett. \textbf{104}, 080501 (2010).

\bibitem{Auyuanet} A. Auyuanet and L. Davidovich. Phys. Rev. A \textbf{82}, 032112 (2010).

\bibitem{Gaussian} Paolo Giorda and Matteo G. A. Paris, Phys. Rev. Lett. \textbf{105}, 020503 (2010).

\bibitem{NP} Yichen Huang, New Journal of Physics \textbf{16}, 033027 (2014).

\bibitem{Rau2009} A. Rau. J. Phys. A:Math. Theor. \textbf{42}, 412002  (2009).
\bibitem{Fedorov} Evgeny Kiktenkoa and Aleksey Fedorov, Phys. Lett. A, \textbf{378} (2014), p. 1704 ;A.K. Fedorov, E.O. Kiktenko, O.V. Man'ko, V.I. Man'ko, arXiv:1409.5265
\bibitem{spinmodels} Raoul Dillenschneider, Phys. Rev. B \textbf{78}, 224413 (2008);  Yichen Huang, Phys. Rev. B \textbf{89}, 054410 (2014).
\bibitem{Werlang} T. Werlang, C. Trippe, G. A. P. Ribeiro, and Gustavo Rigolin, Phys. Rev. Lett. \textbf{105}, 095702 (2010).
\bibitem{Sarandy} M. S. Sarandy, Phys. Rev. A \textbf{80}, 022108 (2009).
\bibitem{XYchain} L. Ciliberti, R. Rossignoli, and N. Canosa, Phys. Rev. A \textbf{82}, 042316 (2010).
\bibitem{swapping} L. Roa, A. Mu\~noz, and G. Gr\"uning, Phys. Rev. A \textbf{89}, 064301 (2014).
\bibitem{Xstates} Qing Chen, Chengjie Zhang, Sixia Yu, X. X. Yi, and C. H. Oh, Phys. Rev. A \textbf{84}, 042313 (2011).

\bibitem{Luo} S. Luo, Phys. Rev. A \textbf{77}, 042303 (2008).

\bibitem{PrimeroX}  M. Ali, A. R. P. Rau, and G. Alber, Phys. Rev. A \textbf{81}, 042105 (2010).

\bibitem{Fanchini}F. F. Fanchini, T. Werlang, C. A. Brasil, L. G. E. Arruda, and A. O. Caldeira, Phys. Rev. A \textbf{81}, 052107 (2010).

\bibitem{Analyticalprogress} Davide Girolami and Gerardo Adesso, Phys. Rev. A \textbf{83}, 052108 (2011).

\bibitem{Optimalmeasurements} Xiao-Ming Lu, Jian Ma, Zhengjun Xi, and Xiaoguang Wang, Phys. Rev. A \textbf{83}, 012327 (2011). 

\bibitem{Proyective} Fernando Galve, Gianluca Giorgi, Roberta Zambrini, Europhys. Lett. \textbf{96} 40005 (2011).

\bibitem{Quesada} N. Quesada, A. Al-Qasimi, and D. F. V. James, J. Mod. Opt. \textbf{59}, 1322 (2012).


\bibitem{Yichen} Yichen Huang, Phys. Rev. A \textbf{88}, 014302 (2013).

\bibitem{Thomas} Thomas M. Cover, Joy A. Thomas. Elements of Information Theory, 2nd Edition (John Wiley \& Sons, Inc. 2006).

\bibitem{Nielsen} Nielsen, M.A., and I. Chuang. Quantum Computation and Quantum Information (Cambridge University Press, Cambridge, England, 2000).

\bibitem{POVM} S. Hamieh, R. Kobes, and H. Zaraket,  Phys. Rev. A \textbf{70}, 052325 (2004) .

\bibitem{Calculo} Stewart, J. Multivariable Calculus, 7th Edition (Cengage Learning 2011).

\bibitem{note} Here the minimization process was performed by Mathematica with the \textit{MinValue} command for given values of $z$.
Wolfram Research, Inc., Mathematica, Version 10.0, Champaign, IL (2010).

\end{thebibliography}
\end{document}